\documentclass[leqno]{article} 

\usepackage{geometry} 
\usepackage{graphicx} 

\usepackage{float} 
\usepackage{wrapfig} 

\usepackage{amsmath}
\usepackage{amssymb}
\usepackage{MnSymbol}
\usepackage{multicol}
\usepackage{hyperref}
\usepackage{tabularx}
\usepackage{qtree}
\usepackage{enumerate}
\usepackage{setspace}
\usepackage{array}
\usepackage{appendix}
\usepackage[all]{xy}

\usepackage{environ}
\makeatletter
\NewEnviron{Lalign}{\tagsleft@true\begin{align}\BODY\end{align}}
\makeatother

\usepackage{environ}
\makeatletter
\NewEnviron{Ralign}{\tagsleft@false\begin{align}\BODY\end{align}}
\makeatother

\graphicspath{{Pictures/}} 

\newcommand{\bsymb}[2]{#1\!:_b\!#2}
\newcommand{\bzero}[1]{\bsymb{#1}{0}}
\newcommand{\bone} [1]{\bsymb{#1}{1}}
\newcommand{\bo}[1]{$\boldsymbol{#1}$}
\newcommand{\nbo}[1]{$#1$}

\newcommand{\Nat}{{\mathbb N}}

\newcommand{\Int}{{\mathbb Z}}

\newcommand{\app}[1]{\hspace{.4mm}{:_{\hspace{.2mm}#1}}\hspace{.4mm}}
\newcommand{\apb}{\app b}
\newcommand{\apbz}{\apb 0}
\newcommand{\apbu}{\apb 1}
\newcommand{\apbo}{\apbu}
\newcommand{\apd}{\app d}

\newcommand{\conca}[1]{\hspace{1mm}\hat{\raisebox{-,4ex}{\footnotesize{\it #1}}}\hspace{1mm}}

\newcommand{\concb}{\conca b}
\newcommand{\concd}{\conca d}

\makeatletter
\renewcommand*{\@fnsymbol}[1]{\ensuremath{\ifcase#1\or \dagger\or \ddagger\or
   \mathsection\or \mathparagraph\or \|\or **\or \dagger\dagger
   \or \ddagger\ddagger \else\@ctrerr\fi}}
\makeatother

\begin{document}


\title{Number representations and term rewriting \thanks{A research project carried out as part of the Honours programme Computer Science and Artificial Intelligence at the University of Amsterdam, under supervision of Alban Ponse and Inge Bethke}}
\author{Boas Kluiving \and Wijnand van Woerkom}
\date{} 

\maketitle

\thispagestyle{empty}

\begin{abstract}
    In this paper we examine a number of term rewriting system for integer number representations, building further upon the datatype defining systems described in \cite{bergstra2014three}. In particular, we look at automated methods for proving confluence and termination in binary and decimal term rewriting systems for both append and tree constructor functions. We find that some of these term rewriting systems are not strongly terminating, which we resolve with minor changes to these systems. Moreover, most of the term rewriting systems discussed do not exhibit the confluence property, which seems more difficult to resolve.
\end{abstract}


\def\bibsection{\section*{References}}

{\setstretch{0.8}
\small \tableofcontents 
}
\newpage


\section{Introduction}

A term rewriting system (TRS) \cite{baader1999term} can be used to represent many different kinds of datatypes and operations. One possibility for which TRSs can be used is for a number representation. Sometimes we will refer to these as datatype defining rewrite systems (DDRSs) \cite{bergstra2014three}. In the case of these DDRSs, term-rewriting rules make it possible to perform addition, negation, subtraction and multiplication on numbers by applying simple rules. There are different ways to represent these rules such that these operations are made possible. The decimal numbers for example can be represented as appended by 10 different unary functions or as a tree of binary functions. In \cite{bergstra2014three} these different possibilities are explored and multiple TRSs for number representations are proposed.

However, for most of these systems no proofs of completeness have yet been presented. In this paper we try to find methods to prove or disprove the completeness of these TRSs and apply them to the systems mentioned in \cite{bergstra2014three}. We started by using manual proof methods but eventually switched to automated ones provided by AProVE \cite{vgiesl2006aprove} and CSI \cite{zankl2011csi}, due to the relatively large size of the systems. An overview of our results can be found in Table \ref{tab:results}. The systems, including small adjustments, can be found in Tables \ref{tab:natbinview} to \ref{DDRSZ}.

We use the same conventions for describing term rewriting systems as in \cite{baader1999term,bergstra2014three,specificatiedictaat} but explain some of it here for ease of reading. The rules in the tables are labeled and some of them have a sub- and superscript; these indicate the rule represents a set of rules (called \textit{rule schemata} \cite{walters1995rewrite}) in which the values of the variables in the subscript are instantiated with natural numbers in the range also denoted by the scripts. E.g. rule [d30.$i.j\hspace{.4mm}]_{i,j=0}^9$ $(-(y\apd j)) + (x\apd i) \rightarrow  P^j((x+(-y))\apd i )$ (Table \ref{tab:intdecview}) represents 100 rules, one for each pair $i, j \in \{0, \dots, 9\}$. In addition the ``digit counter'' variables can be postfixed with $'$ or $^\star$, meaning ``the successor of'' and ``10 minus'' respectively. So informally put $i^\prime$ means $i + 1$ and $i^\star$ means $10 - i$.

\subsection{Ground-completeness}
An important criterion for the usefulness of a TRS is that of ground-completeness. In order to be ground-complete the TRS needs to always have only terminating sequences of rewrites for each term, and every term needs to have its own unique normal form (i.e. it is not possible that a term has two normal forms). These properties are a necessity when trying to construct a DDRS, since we would like every operation to give a result within a finite amount of steps and a term can not have two different final answers depending on which rewriting rules were followed (i.e. produces unique normal forms). A TRS needs to be ground-confluent and terminating in order to be ground-complete. In this paper we are interested in proving these properties of the TRSs of number representations in order to prove the correctness of these systems. 

Ground-confluence however is rather difficult to prove. Although there are some specific methods defined for proving this property \cite{kapur1990ground}, these only work for specific small TRSs. There is no general method for proving ground-confluence. Instead, we try to prove a stronger property than ground-confluence, namely confluence. In order for a TRS to be confluent, the following needs to hold: for every pair of rewrite sequences of the form $s \twoheadrightarrow t_1$, $s \twoheadrightarrow t_2$ there needs to be a term $t$ such that $t_1 \twoheadrightarrow t$ and $t_2 \twoheadrightarrow t$. The two-placed infix function $\twoheadrightarrow$ is defined as follows: for every $t$ $t \twoheadrightarrow t$ and if there exists $u \twoheadrightarrow v$ and $v \rightarrow w$, then $u \twoheadrightarrow w$.

Furthermore, in order for a TRS to be terminating, every possible term needs to be able to be rewritten in a finite amount of steps to a term that cannot be rewritten any further.


\section{Methods}
\subsection{Proof techniques for confluence and termination}
There are several methods for finding proofs for these properties. At the start of this research we made use of two proof techniques that are summarized here for completeness. 

\subsubsection{Knuth-Bendix algorithm}
In order to prove confluence, one possibility is to use the Knuth-Bendix completion algorithm \cite{knuth1983simple}. The Knuth-Bendix algorithm finds those rules that have overlapping redex patterns, meaning that these rules, with the variables partially instantiated, can be matched to the first argument of the $\rightarrow$ operator of two different term rewriting rules. For this term a proof is then needed that these two rewriting rules, after being rewritten, still end up being the same term. If there exists an overlapping redex pattern that cannot be reduced to this same common reduct, confluence is disproven. On the other hand, if these overlapping redex patterns can always be rewritten to the same common reduct, the system is proven to be confluent \cite{knuth1983simple,specificatiedictaat}.

Provided that the TRS in question is terminating, this algorithm can always determine in finite time whether a TRS is confluent, since for every overlapping redex pattern, there are only finitely many steps possible to rewrite the two terms to find if these are rewritable to the same common reduct. Thus, if we can first prove that the TRS is terminating, we can always use the Knuth-Bendix algorithm to prove or disprove confluence.

\subsubsection{Recursive tree orderings}
One method for proving termination is to use recursive tree orderings \cite{dershowitz1979orderings}. This method is defined as follows: \\
$\mathcal{B}$ is defined as the collection of finite commutative trees, with the nodes being labeled with natural numbers. Let $\mathcal{B}^*$ be the collection of such trees as $\mathcal{B}$, but with some of the nodes labeled with $*$ in addition to its natural number.
On $\mathcal{B}^*$ the following reduction-relation is defined: 
\\[2mm]
Let $\vec{t} = t_1, \dots ,t_k$.
\begin{enumerate}
    \item $n(\vec{t}) \rightarrow n^*(\vec{t})$
    \item if $n > m$, then $n^*(\vec{t}) \rightarrow m(n^*(\vec{t}), \dots ,n^*(\vec{t}))$
    \item $n^*(s,\vec{t}) \rightarrow n(s^*,\vec{t})$
    \item $n^*(t_1, \dots ,t_k)\rightarrow t_i$ with $1 \le i \le k$ 
    \item these rules may also be carried out in a specific context, e.g. if $s \rightarrow t$ then also $n(\dots, s, \dots) \rightarrow n(\dots, t, \dots)$
\end{enumerate}
This rewriting relation, when used with $\mathcal{B}$ is terminating. To apply this to TRSs, we can allot all function symbols of a TRS a weight $g$, such that $g(f) > g(h)$ if $f(s_1, \dots, s_n) \rightarrow h(t_1,\dots,t_k)$ and such that the rewriting rules become rewriting rules of elements in $\mathcal{B}$. To prove termination of such a rewriting system, it is sufficient to show that these rewriting rules in $\mathcal{B}$ correspond to one or multiple steps rewritings in $\mathcal{B}^*$ \cite{dershowitz1979orderings,specificatiedictaat}. 

In Appendix A we have worked out a proof using these two techniques. Important to note however is that these methods are sufficient but not necessary methods for the respective properties and that for large TRSs (not an uncommon scenario for a DDRS) these methods may take too long to apply.

\subsection{Automating confluence and termination proofs}
To prove confluence and termination for these larger TRSs we looked for methods of automating confluence and termination proofs. For termination proofs we used AProVE \cite{vgiesl2006aprove}, a tool that uses a variety of methods to try to prove the termination property for TRSs. For confluence proofs we used CSI \cite{zankl2011csi}, which also combines many different proof techniques for TRSs in order to try to prove the confluence property. These automatic proofs can be found at \url{https://staff.fnwi.uva.nl/a.ponse/term_rewriting_proofs/}.


\section{Results and discussion}
In the next subsections we will go over the results in Table \ref{tab:results} individually. In some cases adjustments to the original DDRSs were needed. We have included the edited versions in this paper. For a complete description on how these rules are to be read we refer the reader to \cite{bergstra2014three}, or subsequent versions thereof. In contrast to \cite{bergstra2014three} we write $\rightarrow$ instead of the equality sign because we focussed on term rewriting and not the equational theories that these systems are based on. We have also omitted the rules that serve for translating representations to other representations because these do not interfere with confluence or termination. In some cases this omission causes a gap in the numbering of the rules. We allow this gap in an attempt to match the names of the rules to those used in \cite{bergstra2014three}.

\begin{table}[h]
\centering
\begin{tabularx}{\textwidth}{lll}
\textbf{Rewrite system} &\textbf{Confluence} & \textbf{Termination} \\
\hline
Binary append for $\mathbb{N}$ & Manually proven & Automatically proven \\
Edited binary append for $\mathbb{N}$ & Manually proven & Manually proven \\
Binary append for $\mathbb{Z}$ & Automatically disproven & Automatically proven \\
Decimal append for $\mathbb{N}$& Automatically proven & Automatically disproven \\
Edited decimal append for $\mathbb{N}$ & Automatically proven & Automatically proven \\
Edited decimal append for $\mathbb{Z}$ & Manually disproven & Automatically proven \\
Binary tree constructor for $\mathbb{Z} $& Automatically disproven & Automatically proven \\
Decimal tree constructor for $\mathbb{Z}$ & Automatically disproven & Automatically disproven \\
Edited decimal tree constructor for $\Nat$ & Automatically disproven & Automatically proven \\
Edited decimal tree constructor for $\mathbb{Z}$ & Automatically disproven & Undecided \\
Ring specification for $\mathbb{Z}$ & Automatically disproven & Automatically proven \\
\hline
\end{tabularx}
\caption{Overview of results}
\label{tab:results}
\end{table}

\subsection{Binary append}

\begin{table}
\centering
\hrule
\begin{minipage}[t]{0.52\linewidth}\centering
\begin{Lalign}
\label{b1}
\tag*{[b1.$i\hspace{.4mm}]_{i=0}^1$}
0\apb i & \rightarrow i\\[2mm]
\label{b2}
\tag*{[b2]}
S(0) & \rightarrow  1\\
\label{b3}
\tag*{[b3]}
S(1) & \rightarrow  1\apbz\\
\label{b4}
\tag*{[b4]}
S(x\apbz) & \rightarrow x\apbo\\
\label{b5}
\tag*{[b5]}
S(x\apbo) & \rightarrow S(x)\apbz\\[2mm]
\label{b6}
\tag*{[b6]}
x+0 & \rightarrow x\\
\label{b7}
\tag*{[b7]}
0 + x & \rightarrow x\\
\label{b8}
\tag*{[b8]}
x+1 & \rightarrow S(x)\\
\label{b9}
\tag*{[b9]}
1+x & \rightarrow S(x)\\
\label{b10}
\tag*{[b10.$i.j\hspace{.4mm}]_{i,j=0}^1$}
(x\apb i) + (y\apb j) & \rightarrow  S^j((x+y)\apb i)
\end{Lalign}
\end{minipage}
\begin{minipage}[t]{0.45\linewidth}\centering
\begin{Lalign}
\label{b11}
\tag*{[b11]}
x \cdot 0 & \rightarrow 0\\
\label{b12}
\tag*{[b12]}
x \cdot 1 & \rightarrow x\\
\label{b13}
\tag*{[b13.$i\hspace{.4mm}]_{i=0}^1$}
x\cdot(y\apb i) & \rightarrow ((x\cdot y)\apbz )+(x\cdot i)
\end{Lalign}
\end{minipage}
\begin{minipage}[t]{0.39\linewidth}\centering
\begin{Lalign}
\label{b16}
\tag*{[b16]}
-0 & \rightarrow 0\\
\label{b17}
\tag*{[b17]}
-(-x) & \rightarrow  x\\[2mm]
\label{b18}
\tag*{[b18]}
P(0) &  \rightarrow -1\\
\label{b19}
\tag*{[b19]}
P(1) & \rightarrow  0\\
\label{b20}
\tag*{[b20]}
P(x \apbz) & \rightarrow  P(x)\apbu\\
\label{b21}
\tag*{[b21]}
P(x \apbo) & \rightarrow  x\apbz\\
\label{b22}
\tag*{[b22]}
P(-x) &  \rightarrow  -S(x)\\[2mm]
\label{b23}
\tag*{[b23]}
S(-1) & \rightarrow 0\\
\label{b24}
\tag*{[b24]}
S(-( x\apbz)) & \rightarrow  -(P(x) \apbu) \\
\label{b25}
\tag*{[b25]}
S(-(x \apbu)) & \rightarrow  -(x \apbz)
\end{Lalign}
\end{minipage}
\begin{minipage}[t]{0.6\linewidth}\centering
\begin{Lalign}
\label{b26}
\tag*{[b26]}
(-x) \apbz & \rightarrow  -(x \apbz)\\
\label{b27}
\tag*{[b27]}
(-x) \apbu & \rightarrow  -(P(x) \apbu)\\[2mm]
\label{b28}
\tag*{[b28]}
x + (-1) & \rightarrow  P(x)\\
\label{b29}
\tag*{[b29]}
(-1) + x & \rightarrow  P(x)\\
\label{b30}
\tag*{[b30.$i.j\hspace{.4mm}]_{i,j=0}^1$}
(x\apb i) + (-(y\apb j)) & \rightarrow  P^j((x+(-y))\apb i)\\
\label{b31}
\tag*{[b31.$i.j\hspace{.4mm}]_{i,j=0}^1$}
(-(y\apb j)) + (x\apb i) & \rightarrow  P^j((x+(-y))\apb i)\\
\label{b32}
\tag*{[b32]}
(-x) + (-y) & \rightarrow  -(x+y)\\[2mm]
\label{b33}
\tag*{[b33]}
x \cdot (-y) & \rightarrow  -(x \cdot y)
\end{Lalign}
\end{minipage}\vspace{4mm}
\hrule
\caption{A DDRS for~$\Int_{bud}$ that specifies integer numbers in binary view}
\label{tab:natbinview}
\end{table}

In order to become familiar with the theory and proof techniques the first system we examined was the one using binary append functions for $\mathbb{N}$. In general the downside of using the append representation is that it requires one function for each constant. This tends to make the append systems consist of a large number of rules. In base-2 there are two numeric constants so only $:_b \! 0$ and $:_b \! 1$ have to be considered, making this system a good starting point. The proof itself can be found in the appendix.

During work on the termination proof an issue arose with rule [b10] as it is defined in \cite{bergstra2014three}, we provide it here for the readers convenience: $(x \apb i) + (y \apb j) \rightarrow ((x + y) \apb i) + j$. Using the tree ordering representations of this rule we found that $4(2(0), 2(0)) \!\not\twoheadrightarrow 4(2(4(0, 0)), 0)$ which means termination can not be proven using this method. Looking at the original version of [b10] we can see that it is not strictly necessary to use $+ j$ in this case because $j$ only ranges from 0 to 1. So, to resolve this issue whilst maintaining the equational theory we proposed to change [b10.$i.j\hspace{.4mm}]_{i,j=0}^1$ to $(x\apb i) + (y\apb j) \rightarrow S^j((x+y)\apb i)$, where $S^j$ is notation for taking the successor $j$ times. This changes the recursive tree ordering representation to $2(4(0, 0))$, and $4(2(0), 2(0)) \!\twoheadrightarrow 2(4(0, 0))$. With this change the manual termination proof of $\Nat_{bud}$ and $\Int_{bud}$ could be completed (this corresponds with the ``edited" version in Table \ref{tab:results}). Using AProVE we found the system is also terminating without applying the change. 

The confluence proof was done using the redex pattern technique. A few overlapping patterns were discovered but could all be rewritten to a common reduction. This means the edited TRS for $\Nat_{bud}$ is also confluent and therefore complete. See the appendix for more details on these manual proofs. 

After examining the termination proof for the original binary append TRS for $\mathbb{N}$ by AProVE, it appears that this TRS is terminating as well. Although it was not possible to use the recursive tree ordering technique for proving termination for the TRS, its termination could still be proven using other automated techniques.

For the $\mathbb{Z}$ versions of the binary append TRS (see Table \ref{tab:natbinview}), termination could be proven in the same way using AProVE. However, confluence could be disproven using the following rewritings
\begin{displaymath}
    \xymatrix{
        & P(-(-x)) \ar@{->}[dl]_{\text{[b17]}}\ar@{->}[dr]^{\text{[b22]}}\\
        P(x) & & -S(-x)
    }
\end{displaymath}
\noindent because $P(x) \not\twoheadrightarrow -S(-x)$ and $-S(-x) \not\twoheadrightarrow P(x)$. Also, Knuth-Bendix completion was tried in this case to find out whether this system can be made confluent with the addition of a few rules. This however was not very well possible, since the addition of the new rewriting rules would then interfere with the earlier rules, making it necessary to change the existing rules to be able to make the system confluent. 

This however does not exclude the possibility that the TRS itself is confluent for ground terms only (ground-confluent). Unfortunately this is a lot harder to prove and for this problem we could not find general approaches.

\subsection{Decimal append}

\begin{table}
\centering
\hrule
\begin{minipage}[t]{0.51\linewidth}\centering
\begin{Lalign}
\label{d1}
\tag*{[d1.$i\hspace{.4mm}]_{i=0}^9$}
0 \apd i & \rightarrow  i\\[2mm]
\label{d2}
\tag*{[d2.$i\hspace{.4mm}]_{i=0}^8$}
S(i) & \rightarrow i^{\prime}\\
\label{d3}
\tag*{[d3]}
S(9) & \rightarrow  1 \apd 0\\
\label{d4}
\tag*{[d4.$i\hspace{.4mm}]_{i=0}^8$}
S(x \apd i) & \rightarrow  x \apd i^{\prime}\\
\label{d5}
\tag*{[d5]}
S(x \apd 9) & \rightarrow  S(x) \apd 0\\[2mm]
\label{d6}
\tag*{[d6]}
x + 0 & \rightarrow x\\
\label{d7}
\tag*{[d7]}
0 + x & \rightarrow x\\
\label{d8}
\tag*{[d8.$i\hspace{.4mm}]_{i=1}^9$}
x+i & \rightarrow  S^i(x)\\
\label{d9}
\tag*{[d9.$i\hspace{.4mm}]_{i=1}^9$}
i + x & \rightarrow  S^i(x)\\
\label{d10}
\tag*{[d10.$i.j\hspace{.4mm}]_{i,j=0}^9$}
(x \apd i)+(y \apd j) & \rightarrow  S^j((x +y) \apd i)
\end{Lalign}
\end{minipage}\vspace{4mm}
\hfill
\begin{minipage}[t]{0.48\linewidth}\centering
\begin{Lalign}
\label{d11}
\tag*{[d11]}
x \cdot 0 & \rightarrow 0\\
\label{d12}
\tag*{[d12.$i\hspace{.4mm}]_{i=0}^8$}
x \cdot i^{\prime}  &  \rightarrow  (x \cdot i) + x\\
\label{d13}
\tag*{[d13.$i\hspace{.4mm}]_{i=0}^9$}
x \cdot (y \apd i) & \rightarrow  ((x \cdot y) \apd 0) + (x \cdot i)
\end{Lalign}
\end{minipage}
\begin{minipage}[t]{0.4\linewidth}\centering
\begin{Lalign}
\label{d15}
\tag*{[d15]}
-0 & \rightarrow 0\\
\label{d16}
\tag*{[d16]}
-(-x) & \rightarrow  x\\[2mm]
\label{d17}
\tag*{[d17]}
P(0)& \rightarrow -1\\
\label{d18}
\tag*{[d18.$i\hspace{.4mm}]_{i=0}^8$}
P(i^\prime)  & \rightarrow   i\\
\label{d19}
\tag*{[d19]}
P(x \apd 0) & \rightarrow  P(x) \apd 9\\
\label{d20}
\tag*{[d20.$i\hspace{.4mm}]_{i=0}^8$}
P(x \apd i^\prime) & \rightarrow  x \apd i\\
\label{d21}
\tag*{[d21]}
P(-x) & \rightarrow  -S(x)
\end{Lalign}
\end{minipage}
\hfill
\begin{minipage}[t]{0.58\linewidth}\centering
\begin{Lalign}
\label{d22}
\tag*{[d22.$i\hspace{.4mm}]_{i=0}^8$}
S(-i^\prime)  & \rightarrow  - i\\
\label{d23}
\tag*{[d23]}
S(-(x \apd 0)) & \rightarrow  -(P(x) \apd 9)\\
\label{d24}
\tag*{[d24.$i\hspace{.4mm}]_{i=0}^8$}
S(-(x \apd i^\prime)) & \rightarrow  -(x \apd i)\\[2mm]
\label{d25}
\tag*{[d25]}
(-x) \apd 0 & \rightarrow  -(x \apd 0)\\[0mm]
\label{d26}
\tag*{[d26.$i\hspace{.4mm}]_{i=1}^9$}
(-x) \apd i & \rightarrow  -(P(x) \apd i^{\star})
\end{Lalign}
\end{minipage}
\begin{minipage}[t]{1\linewidth}\centering
\begin{Lalign}
\label{d27}
\tag*{[d27.$i\hspace{.4mm}]_{i=1}^9$}
x + (- i) & \rightarrow  P^i(x)\\
\label{d28}
\tag*{[d28.$i\hspace{.4mm}]_{i=1}^9$}
(- i) + x & \rightarrow  P^i(x)\\
\label{d29}
\tag*{[d29.$i.j\hspace{.4mm}]_{i,j=0}^9$}
(x\apd i) + (-(y\apd j)) & \rightarrow  P^j((x+(-y))\apd i )\\
\label{d30}
\tag*{[d30.$i.j\hspace{.4mm}]_{i,j=0}^9$}
(-(y\apd j)) + (x\apd i) & \rightarrow  P^j((x+(-y))\apd i )\\
\label{d31}
\tag*{[d31]}
(-x) + (-y) & \rightarrow  -(x+y)\\[2mm]
\label{d32}
\tag*{[d32]}
x \cdot (-y) & \rightarrow  -(x \cdot y)
\end{Lalign}
\end{minipage}\vspace{4mm}
\hrule
\caption{A DDRS for~$\Int_{dub}$ that specifies integers in decimal view}
\label{tab:decview}
\end{table}

In base 10 we have append functions $\apd 0, \dots, \apd 9$. Even though the nature of the rules is very similar to those in the binary system we have many more constants and append functions. This means that rules such as [d10.$i.j\hspace{.4mm}]_{i,j=0}^9$ (see Table \ref{tab:decview}) make the TRS grow exponentially in size. The assistance of automated theorem provers such as CSI and AProVE becomes more and more useful for bigger systems such as this one.

Using AProVE to test whether this system $Z_{dub}$ as described in \cite{bergstra2014three} is terminating, we found the following rewriting loops
\[
\begin{array}{rl@{\hspace{2em}}l}
1 + 0 \!\!\!\!\! & \rightarrow S(0) + 0 & \text{[d9.0]} \\
& \rightarrow 1 + 0 & \text{[d2.0],}\\[2mm]
-1 + 0 \!\!\!\!\! & \rightarrow P(0) + -0 & \text{[d28.0]} \\
& \rightarrow -1 + -0 & \text{[d17]} \\
& \rightarrow -1 + 0 & \text{[d15].}
\end{array}
\]

\noindent Thus we can conclude that the decimal append TRS in \cite{bergstra2014three} is not strongly terminating and a change in the term rewriting system is needed to acquire the termination property. The loops are caused mainly by the variant of the rule we proposed to change in the binary system. Although this change was unrelated to the change of the binary append rewriting system, a similar change in this system resolves the loop. So, we proposed to change [d9.$i\hspace{.4mm}]_{i=1}^9$ to $i+ x \rightarrow S^i(x)$ (and similarly d8 and d10), and [d28.$i\hspace{.4mm}]_{i=1}^9$ to $(- i) + x \rightarrow P^i(x)$ (and similarly d27, d29, and d30). This results in the DDRS described in Table \ref{tab:decview}. Applying these changes, termination could be proven for the $\Int_{dub}$ TRS using AProVE.

For $\mathbb{N}$ and $\mathbb{Z}$, these changes were sufficient to prove termination. For $\mathbb{N}$ confluence could also be proven automatically by CSI. Unfortunately for $\mathbb{Z}$ the system was too large to prove confluence with CSI. However, the same counterexample that was found for binary append is applicable for the decimal append term rewriting system:

\begin{displaymath}
    \xymatrix{
        & P(-(-x)) \ar@{->}[dl]_{\text{[d16]}}\ar@{->}[dr]^{\text{[d21]}}\\
        P(x) & & -S(-x)
    }
\end{displaymath}

\noindent where $P(x) \not\twoheadrightarrow -S(-x)$ and $-S(-x) \not\twoheadrightarrow P(x)$. Thus, the decimal append system for $\mathbb{Z}$ is non-confluent as well.

\subsection{Binary tree constructor}

\begin{table}
\hrule
\begin{minipage}[t]{0.4\linewidth}\centering
\begin{Lalign}
\label{bi1}
\tag*{[bt1]}
0\concb x& \rightarrow x\\
\label{bi2}
\tag*{[bt2]}
x\concb(y\concb z)& \rightarrow (x+ y)\concb z\\[2mm]
\label{bi3}
\tag*{[bt3]}
0+x & \rightarrow x\\
\label{bi4}
\tag*{[bt4]}
x+0 & \rightarrow x\\
\label{bi5}
\tag*{[bt5]}
1+1 & \rightarrow 1\concb 0\\
\label{bi6}
\tag*{[bt6]}
x + (y\concb z) & \rightarrow  y\concb(x+z)\\
\label{bi7}
\tag*{[bt7]}
(x \concb y)+z & \rightarrow  x\concb(y+z)
\\[2mm]
\label{bi8}
\tag*{[bt8]}
x \cdot 0 & \rightarrow 0\\
\label{bi9}
\tag*{[bt9]}
0\cdot x& \rightarrow 0\\
\label{bi10}
\tag*{[bt10]}
1\cdot 1& \rightarrow 1\\
\label{bi11}
\tag*{[bt11]}
x \cdot (y\concb z) & \rightarrow (x \cdot y)\concb(x\cdot z)\\
\label{bi12}
\tag*{[bt12]}
(x \concb y)\cdot z & \rightarrow (x \cdot z)\concb(y\cdot z)
\end{Lalign}
\end{minipage}\vspace{4mm}
\hfill
\begin{minipage}[t]{0.5\linewidth}\centering
\begin{Lalign}
\label{bi13}
\tag*{[bt13]}
-0& \rightarrow 0\\
\label{bi14}
\tag*{[bt14]}
-(-x)& \rightarrow x\\[2mm]
\label{bi15}
\tag*{[bt15]}
1\concb (-1)& \rightarrow 1\\
\label{bi16}
\tag*{[bt16]}
(x\concb 0)\concb (-1)& \rightarrow (x\concb(-1))\concb1\\
\label{bi17}
\tag*{[bt17]}
(x\concb 1)\concb (-1)& \rightarrow (x\concb0)\concb 1\\
\label{bi18}
\tag*{[bt18]}
x\concb (-(y\concb z))& \rightarrow -((y+(-x))\concb z)\\
\label{bi19}
\tag*{[bt19]}
(-x)\concb y& \rightarrow -(x\concb(-y))\\[2mm]
\label{bi20}
\tag*{[bt20]}
1+(-1)& \rightarrow 0\\
\label{bi21}
\tag*{[bt21]}
(-1)+1& \rightarrow 0\\
\label{bi22}
\tag*{[bt22]}
(-1)+(-1)& \rightarrow -(1\concb 0)\\
\label{bi23}
\tag*{[bt23]}
x+(-(y\concb z))& \rightarrow -(y\concb (z+(-x)))\\
\label{bi24}
\tag*{[bt24]}
(-(x\concb y))+z& \rightarrow -(x\concb (y+(-z)))\\[2mm]
\label{bi25}
\tag*{[bt25]}
x\cdot (-y)& \rightarrow -(x\cdot y)\\
\label{bi26}
\tag*{[bt26]}
(-x)\cdot y& \rightarrow -(x\cdot y)
\end{Lalign}
\end{minipage}
\hrule
\caption{A DDRS for~$\Int_{bt}$, integer numbers in binary view with binary digit tree constructor}
\label{natbview}
\end{table}

The tree constructor functions $\concb$ and $\concd$ take two arguments. This reduces the number of functions and therefore the number of rules in the TRS. The binary system consists of 26 rules and does not contain the loop found in the decimal append system (see Table \ref{natbview}). The TRS for $\Nat_{bt}$ was taken from- and proven to be confluent and terminating in \cite{BW89}, so we considered $\Int_{bi}$ as defined in \cite{bergstra2014three}. AProVE could prove termination for this system as well.

When proving confluence for the binary tree constructor term rewriting system a counterexample was found:
\begin{displaymath}
    \xymatrix{
        & x \concb (y \concb (z \concb w)) \ar@{->>}[dl]\ar@{->>}[dr]\\
        (x + (y + z)) \concb w & & ((x + y) + z) \concb w
    }
\end{displaymath}
because $(x + (y + z))\concb w \not \twoheadrightarrow ((x + y) + z)\concb w$ and $ ((x + y) + z)\concb w \not \twoheadrightarrow (x + (y + z))\concb w$. While we can conclude from this example that the term rewriting system is indeed non-confluent, it could still be ground-confluent. The finding that this TRS is not confluent is however contrary to that in \cite{BW89}, who claim that this system is confluent and terminating. In subsequent papers, this claim was reduced to ground-confluence. 

\subsection{Decimal tree constructor}

\begin{table}
\hrule
\begin{minipage}[t]{0.46\linewidth}\centering
\begin{Lalign}
\label{dt1}
\tag*{[dt1]}
0\concd x& \rightarrow x\\
\label{dt2}
\tag*{[dt2]}
x\concd (y\concd z)& \rightarrow (x+y)\concd z\\[2mm]
\label{dt3}
\tag*{[dt3.$i\hspace{.4mm}]_{i=0}^8$}
S(i)& \rightarrow i'\\
\label{dt4}
\tag*{[dt4]}
S(9)& \rightarrow 1\concd 0\\
\label{dt5}
\tag*{[dt5.$i\hspace{.4mm}]_{i=0}^8$}
S(x\concd i)& \rightarrow x\concd i'\\
\label{dt6}
\tag*{[dt6]}
S(x\concd 9)& \rightarrow S(x)\concd 0\\[2mm]
\label{dt7}
\tag*{[dt7]}
x+0 & \rightarrow x\\
\label{dt8}
\tag*{[dt8.$i\hspace{.4mm}]_{i=0}^8$}
x+i'& \rightarrow S(x)+i\\
\label{dt9}
\tag*{[dt9.$i\hspace{.4mm}]_{i=0}^9$}
x+(y\concd i)& \rightarrow (y\concd x)+i\\[2mm]
\label{dt10}
\tag*{[dt10]}
x\cdot 0 & \rightarrow 0\\
\label{dt11}
\tag*{[dt11.$i\hspace{.4mm}]_{i=0}^8$}
x\cdot i'& \rightarrow x+(x\cdot i)\\
\label{dt12}
\tag*{[dt12.$i\hspace{.4mm}]_{i=0}^9$}
x\cdot (y\concd i)& \rightarrow ((x\cdot y)\concd 0)+(x\cdot i)
\end{Lalign}
\end{minipage}\vspace{4mm}
\hfill
\begin{minipage}[t]{0.5\linewidth}\centering
\begin{Lalign}
\label{dt13}
\tag*{[dt13]}
-0& \rightarrow 0\\
\label{dt14}
\tag*{[dt14]}
-(-x)& \rightarrow x\\[2mm]
\label{dt15}
\tag*{[dt15]}
P(0)& \rightarrow -1\\
\label{dt16}
\tag*{[dt16.$i\hspace{.4mm}]_{i=0}^8$}
P(i^\prime)  & \rightarrow   i\\
\label{dt17}
\tag*{[dt17]}
P(x \concd 0) & \rightarrow  P(x) \concd 9\\
\label{dt18}
\tag*{[dt18.$i\hspace{.4mm}]_{i=0}^8$}
P(x \concd i^\prime) & \rightarrow  x \concd i\\
\label{dt19}
\tag*{[dt19]}
P(-x) & \rightarrow  -S(x)
\\[2mm]
\label{dt20}
\tag*{[dt20.$i\hspace{.4mm}]_{i=0}^8$}
S(-i^\prime)  & \rightarrow  - i\\
\label{dt21}
\tag*{[dt21]}
S(-(x \concd 0)) & \rightarrow  -(P(x) \concd 9)\\
\label{dt22}
\tag*{[dt22.$i\hspace{.4mm}]_{i=0}^8$}
S(-(x \concd i^\prime)) & \rightarrow  -(x \concd i)\\[2mm]
\label{dt23}
\tag*{[dt23]}
(-x)\concd y& \rightarrow -(x\concd(-y))\\[2mm]
\label{dt24}
\tag*{[dt24$.i.j\hspace{.4mm}]_{i,j=1}^9$}
i\concd (-j)& \rightarrow P(i)\concd j^\star\\
\label{dt25}
\tag*{[dt25.$i\hspace{.4mm}]_{i=1}^9$}
(x\concd y)\concd (-i)& \rightarrow P(x\concd y)\concd i^\star\\
\label{dt26}
\tag*{[dt26]}
x\concd (-(y\concd z))& \rightarrow -((y+(-x))\concd z)\\[2mm]
\label{dt27}
\tag*{[dt27.$i\hspace{.4mm}]_{i=1}^9$}
x+(-i)& \rightarrow P^i(x)\\
\label{dt28}
\tag*{[dt28]}
x+(-(y\concd z))& \rightarrow -(y\concd (z+(-x)))\\
\label{dt29}
\tag*{[dt29.$i\hspace{.4mm}]_{i=1}^9$}
(-i)+x& \rightarrow P^i(x)\\
\label{dt30}
\tag*{[dt30]}
(-(x\concd y))+z& \rightarrow -(x\concd (y+(-z)))\\[2mm]
\label{dt31}
\tag*{[dt31]}
x\cdot (-y)& \rightarrow -(x\cdot y)\\
\label{dt32}
\tag*{[dt32]}
(-x)\cdot y& \rightarrow -(x\cdot y)
\end{Lalign}
\end{minipage}
\hrule
\caption{A DDRS for~$\Int_{dt}$, integer numbers with decimal digit tree constructor in 
decimal view}
\label{tab:intdecview}
\end{table}

Similarly to the difference in size between the append function systems for binary and decimal the TRS for $\Int_{dt}$ is considerably larger than that of $\Int_{bi}$. The TRS for $\Nat_{dt}$ in \cite{bergstra2014three} does not contain a similar rewriting loop either because only rule [dt8.$i\hspace{.4mm}]_{i=0}^8$ is present. Moreover AProVE could prove termination for this system.

In \cite{bergstra2014three}, the rules included for the negative numbers in $\Int_{dt}$ do facilitate a similar rewriting loop:
\[
\begin{array}{rl@{\hspace{2em}}l}
-1 + 0 \!\!\!\!\! & \rightarrow P(0) + -0 & \text{[dt29.0]} \\
& \rightarrow -1 + -0 & \text{[dt15]} \\
& \rightarrow -1 + 0 & \text{[dt13]} \\
\end{array}
\]
\noindent so we proposed the same change, leading to [dt27.$i\hspace{.4mm}]_{i=1}^9$ and [dt29.$i\hspace{.4mm}]_{i=1}^9$ (see Table \ref{tab:intdecview}). In contrast to the other systems AProVE was not able to prove or disprove termination after resolving the loop. A potential cause of this is the number of defined symbols used in the TRS (Carsten Fuhs (AProVE contributor), email communication, December 2015). A symbol, $f$, is defined if the TRS contains a rule of the form $f(\dots) \rightarrow \dots$. Symbols for which this is not so are called constructor symbols. None of the TRSs contain any constructor symbols which makes the proof more complex. 

For the confluence proof, CSI was able to find a counterexample for the decimal tree constructor
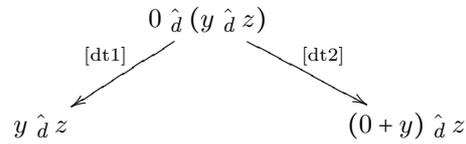
\begin{displaymath}
    \xymatrix{
        & 0 \concd (y \concd z) \ar@{->}[dl]_{\text{[dt1]}}\ar@{->}[dr]^{\text{[dt2]}}\\
        y \concd z & & (0 + y) \concd z
    }
\end{displaymath}
because $y \concd z \not\twoheadrightarrow (0 + y) \concd z$ or $ (0 + y) \concd z \not\twoheadrightarrow y \concd z$. Since this counterexample is valid for both the unedited and the edited version of the decimal tree constructors, we can conclude that these term rewriting systems are not confluent.

\begin{table}
\hrule
\begin{minipage}[t]{0.5\linewidth}\centering
\begin{Lalign}
\label{eq:3}\tag*{[r1]}
-0&\rightarrow0
\\
\label{eq:4}\tag*{[r2]}
-(-x) &\rightarrow x
\\[4mm]
\label{eq:5}\tag*{[r3]}
x+(y+z) &\rightarrow (x + y) + z
\\
\label{eq:6}\tag*{[r4]}
x+0&\rightarrow x\\
\label{eq:7}\tag*{[r5]}
1+(-1)  &\rightarrow 0 \\
\label{eq:8}\tag*{[r6]}
(x + 1)+(-1)  &\rightarrow x \\
\label{eq:9}\tag*{[r7]}
x+(-(y+1))&\rightarrow(x+(-y))+(-1)
\end{Lalign}
\end{minipage}\vspace{4mm}
\hfill
\begin{minipage}[t]{0.47\linewidth}\centering
\begin{Lalign}
\label{eq:10}\tag*{[r8]}
	0+x     &\rightarrow x\\
\label{eq:11}\tag*{[r9]}
	(-1)+1  &\rightarrow 0 \\
\label{eq:12}\tag*{[r10]}
	(-(x+1)) + 1 &\rightarrow -x\\
\label{eq:13}\tag*{[r11]}
	(-x) + (-y) &\rightarrow -(x+y)\\[2mm]
\label{eq:14}\tag*{[r12]}
	x \cdot 0 &\rightarrow 0  \\
\label{eq:15}\tag*{[r13]}
	x\cdot 1 &\rightarrow x \\
\label{eq:16}\tag*{[r14]}
	x\cdot (- y) &\rightarrow (-x)\cdot y \\
\label{eq:17}\tag*{[r15]}
	x\cdot(y + z) &\rightarrow (x\cdot y) + (x\cdot z)
\end{Lalign}
\end{minipage}
\hrule
\caption{A DDRS for $\Int_r$}
\label{DDRSZ}
\end{table}

\subsection{Ring system}
For the ring TRS (Table \ref{DDRSZ}) defined in \cite{bergstra2014fracpairs}, termination could be proven using AProVE. However, the system is not confluent
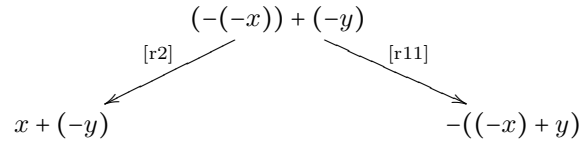
\begin{displaymath}
    \xymatrix{
        & (-(-x)) + (-y) \ar@{->}[dl]_{\text{[r2]}}\ar@{->}[dr]^{\text{[r11]}}\\
        x + (-y) & & -((-x) + y)
    }
\end{displaymath}
because $(-(-x)) + (-y) \not\twoheadrightarrow -((-x) + y)$ and $-((-x) + y) \not\twoheadrightarrow (-(-x)) + (-y)$. Ground-confluence on the other hand is not ruled out in this case either. 

To discover whether this term rewriting system can be made confluent with the addition of a few rules, we tried to use Knuth-Bendix completion. However, by adding new rules in either way, these new rules again conflict with original rules, making it impossible to create a coherent confluent system based on these rules. Too many rules needed to be added and changed to solve the confluence issues. The system itself does not seem to be designed with confluence in mind. 


\section{Conclusion}

Some of the DDRSs in \cite{bergstra2014three} were shown to be both confluent and terminating and therefore ground-complete. Others were disproven and in some cases modified in order to obtain a proof.
We found that to prove whether a TRS is ground-complete (in particular large ones) or not, automated theorem provers such as CSI and AProVE can be useful. 

Most of the DDRSs that were proven not to be confluent are not easy to amend to obtain a confluent TRS using Knuth-Bendix completion. It seems that these systems have some inherent rules that are not suitable for confluence. To ensure confluence, the TRS may need to be designed in another way. Perhaps this could be investigated further to create DDRSs for $\mathbb{Z}$ that are confluent.

One thing to note however is that while we can prove confluence relatively easily, ground-confluence is a lot more difficult to prove. There are no general methods or programs that could be used to prove this property. It would be interesting to perform more research in proving confluence for ground terms and use this to prove ground-confluence for the term rewriting systems above that were proven to be non-confluent. In this case, ground-completeness of those systems could still be proven.

A possible expansion of this work would be to extend the DDRSs to include the modulo and exponentiation operators. 



\begin{appendices}

\section{Appendix}
The automatic proofs can be found at \url{https://staff.fnwi.uva.nl/a.ponse/term_rewriting_proofs/}. For completeness, in the appendices below we included the manual proofs.

\subsection{Termination proof binary append notation} 

Translating the TRS to recursive tree orderings with weights chosen so that \\ $0, 1 < \ \bone{ },\  \bzero{ } < S < + < \cdot $ gives the tree representation displayed in Table \ref{tab:rec_tree}.

\begin{table}[H]
    \centering
        \begin{tabularx}{\textwidth}{c||l@{\extracolsep{1cm}}r@{\extracolsep{0cm}}@{$\ \rightarrow \ $}l}
        \multicolumn{1}{c}{\#} & \multicolumn{1}{l}{Rewriting rule(s)} & \multicolumn{2}{c}{Tree representation} \\
        \hline 
        1 & {[b1.\emph{i}]}$_{i=0}^1$ & 2(0) & 0 \\
        2 & {[b2]} & 3(0) & 0 \\
        3 & {[b3]} & 3(0) & 2(0) \\
        4 & {[b4]} & 3(2(0)) & 2(0) \\
        5 & {[b5]} & 3(2(0)) & 2(3(0)) \\
        6 & {[b6]} \& {[b7]} & 4(0,0) & 0 \\
        7 & {[b8]} \& {[b9]} & 4(0,0) & 3(0) \\
        8 & {[b10.\emph{i}.0]}$_{i=0}^1$ \& {[b10.0.1]} & 4(2(0), 2(0)) & 2(4(0,0)) \\
        9 & {[b10.1.1]} & 4(2(0), 2(0)) & 2(3(4(0,0))) \\
        10 & {[b11]} \& {[b12]} & 5(0,0) & 0 \\
        11 & {[b13.0]} & 5(0,2(0)) & 2(5(0,0)) \\
        12 & {[b13.1]} & 5(0,2(0)) & 4(2(5(0,0)),0) \\
        \hline
    \end{tabularx}
    \caption{Tree representation of $\Nat_{bud}$}
    \label{tab:rec_tree}
\end{table}

Using the rules from Section 2.21 to rewrite the trees, the following can be extracted:
\begin{align}
2(0) & \rightarrow 2^*(0) \rightarrow 0 \\
3(0) & \rightarrow 3^*(0) \rightarrow 0 \\
3(0) & \rightarrow 3^*(0) \rightarrow 2(3^*(0)) \rightarrow 2(0) \\
3(2(0)) & \rightarrow 3^*(2(0)) \rightarrow 2(0) \\
3(2(0)) & \rightarrow 3^*(2(0)) \rightarrow 2(3^*(2(0))) \rightarrow 2(3(2^*(0))) \rightarrow 2(3(0)) \\
4(0,0) & \rightarrow 4^*(0,0) \rightarrow 0 \\
4(0,0) & \rightarrow 4^*(0,0) \rightarrow 3(4^*(0,0)) \rightarrow 3(0) \\
4(2(0), 2(0)) & \rightarrow 4^*(2(0), 2(0)) \rightarrow 2(4^*(2(0), 2(0))) \twoheadrightarrow 2(4(0,0)) \\
4(2(0), 2(0)) & \rightarrow 4^*(2(0), 2(0)) \rightarrow 2(4^*(2(0), 2(0))) \rightarrow 2(3(4^*(2(0), 2(0)))) \\ 
\nonumber & \twoheadrightarrow 2(3(4(0,0))) \\
5(0,0) & \rightarrow 5^*(0,0) \rightarrow 0 \\
5(0,2(0)) & \rightarrow 2(5^*(0,2(0))) \twoheadrightarrow 2(5(0,0)) \\
5(0,2(0)) & \rightarrow 4(5^*(0,2(0)), 5^*(0,2(0))) \twoheadrightarrow 4(5^*(0,2(0)), 0) \\
\nonumber & \rightarrow 4(2(5^*(0,2(0))), 0) \twoheadrightarrow 4(2(5(0,0)), 0) 
\end{align}
For the $\mathbb{Z}_{bud}$ TRS we obtained the following:
\begin{align}
1(0) & \rightarrow 1^*(0) \rightarrow 0 \\
1(1(0)) & \rightarrow 1^*(1(0)) \rightarrow 1(0) \rightarrow 1^*(0) \rightarrow 0 \\
3(0) & \rightarrow 3^*(0) \rightarrow 1(3^*(0)) \rightarrow 1(0) \\
3(0) & \rightarrow 3^*(0) \rightarrow 0 \\
3(2(0)) & \rightarrow 3^*(2(0)) \rightarrow 2(3^*(2(0))) \twoheadrightarrow 2(3(0)) \\
3(2(0)) & \rightarrow 3^*(2(0)) \rightarrow 2(0) \\
3(1(0)) & \rightarrow 3^*(1(0)) \rightarrow 1(3^*(1(0))) \twoheadrightarrow 1(3(0)) \\
3(1(0)) & \rightarrow 3^*(1(0)) \rightarrow 1(0) \rightarrow 1^*(0) \rightarrow 0 \\
3(1(2(0))) & \rightarrow 3^*(1(2(0))) \rightarrow 1(3^*(1(2(0)))) \rightarrow 1(2(3^*(1(2(0))))) \\
\nonumber & \twoheadrightarrow 1(2(3(0))) \\
3(1(2(0))) & \rightarrow 3^*(1(2(0))) \rightarrow 1(3^*(1(2(0)))) \rightarrow 1(2(3^*(1(2(0))))) \\
\nonumber & \twoheadrightarrow 1(2(3(0))) \rightarrow 1(2(3^*(0))) \rightarrow1(2(0)) \\
4(0, 1(0)) & \rightarrow 4^*(0, 1(0)) \rightarrow 3(4^*(0, 1(0))) \rightarrow 3(0) \\ 
4(1(0), 0) & \twoheadrightarrow 3(0), \text{follows from commutativity of the trees and (23)} \\
4(2(0), 1(2(0))) & \rightarrow 4^*(2(0), 1(2(0))) \rightarrow 2(4^*(2(0), 1(2(0)))) \\
\nonumber & \rightarrow 2(4(2^*(0), 1(2(0)))) \rightarrow 2(4(0, 1(2(0)))) \\
\nonumber & \rightarrow 2(4(0, 1^*(2(0)))) \twoheadrightarrow 2(4(0, 1(0)))
\end{align}
\begin{align}
4(2(0), 1(2(0))) & \rightarrow 4^*(2(0), 1(2(0))) \\
\nonumber & \rightarrow 2(4^*(2(0), 1(2(0)))) \\
\nonumber & \rightarrow 2(3(4^*(2(0), 1(2(0))))) \\
\nonumber & \twoheadrightarrow 2(3(4(0, 1(0)))) \\
4(1(0), 1(0)) & \rightarrow 4^*(1(0), 1(0)) \rightarrow 1(4^*(1(0), 1(0))) \\
\nonumber & \twoheadrightarrow 1(4(0,0)) \\
5(0, 1(0)) & \rightarrow 5^*(0, 1(0)) \rightarrow 1(5^*(0, 1(0))) \rightarrow 1(5(0, 1^*(0))) \\
\nonumber & \rightarrow 1(5(0, 0))
\end{align}

\subsection{Confluence proof binary append notation}

To prove confluence we check if there are overlapping redex patterns and if so, if they have a common reduction via rewriting. In order to do this we first write the redex pattern for each rewriting rule: 

\begin{multicols}{3}
\begin{itemize}
\item[(1.0)]
\Tree 
[.\bo{\bzero{}} [.\bo{0} ] ]
\item[(1.1)]
\Tree 
[.\bo{\bone{}} [.\bo{0} ] ]
\item[(2)]
\Tree
[.\bo{S} [.\bo{0} ] ]
\item[(3)]
\Tree
[.\bo{S} [.\bo{1} ] ]
\item[(4)]
\Tree
[.\bo{S} [.\bo{\bzero{}} [.\nbo{x} ] ] ]
\item[(5)]
\Tree
[.\bo{S} [.\bo{\bone{}} [.\nbo{x} ] ] ]
\item[(6)]
\Tree
[.\bo{+} [\nbo{x} \bo{0} ] ]
\item[(7)]
\Tree
[.\bo{+} [\bo{0} \nbo{x} ] ]
\item[(8)]
\Tree
[.\bo{+} [\nbo{x} \bo{1} ] ]
\item[(9)]
\Tree
[.\bo{+} [\bo{1} \nbo{x} ] ]
\item[(10.0.0)]
\Tree
[.\bo{+} [[.\bo{\bzero{}} [.\nbo{x} ] ] [.\bo{\bzero{}} [.\nbo{y} ] ] ] ]
\item[(10.0.1)]
\Tree
[.\bo{+} [[.\bo{\bzero{}} [.\nbo{x} ] ] [.\bo{\bone{}} [.\nbo{y} ] ] ] ]
\item[(10.1.0)]
\Tree
[.\bo{+} [[.\bo{\bone{}} [.\nbo{x} ] ] [.\bo{\bzero{}} [.\nbo{y} ] ] ] ]
\item[(10.1.1)]
\Tree
[.\bo{+} [[.\bo{\bone{}} [.\nbo{x} ] ] [.\bo{\bone{}} [.\nbo{y} ] ] ] ]
\item[(11)]
\Tree
[.\bo{\cdot} [\nbo{x} \bo{0} ] ]
\item[(12)]
\Tree
[.\bo{\cdot} [\nbo{x} \bo{1} ] ]
\item[(13)]
\Tree 
[.\bo{\cdot} [\nbo{x} [.\bo{\bzero{}} \nbo{y} ] ] ]
\item[(14)]
\Tree
[.\bo{\cdot} [\nbo{x} [.\bo{\bone{}} \nbo{y} ] ] ] 
\end{itemize}
\end{multicols}
\newpage
Some overlap exists between these redex patterns. To prove confluence we show that they can be rewritten to a common reduction.
\begin{enumerate}[i.]
\item (4) with (1.0): $S(0\bzero{})$ \\
$S(0\bzero{}) \rightarrow_{1.0} S(0) \rightarrow_{2} 1 $ \\
$S(0\bzero{}) \rightarrow_{4} \bone{0} \rightarrow_{1.1} 1$ \\
So both these terms can be rewritten to a common term.

\item (5) with (1.1): $S(0\bone{})$ \\
$S(0\bone{}) \rightarrow_{1.1} S(1) \rightarrow_{3} \bzero{1}$ \\
$S(0\bone{}) \rightarrow_{5} S(0)\bzero{} \rightarrow_{2} \bzero{1}$ \\
So both these terms can be rewritten to a common term.
\item (6) with (7): $0 + 0$ \\
$0 + 0 \rightarrow_{6} 0$ \\
$0 + 0 \rightarrow_{7} 0$ \\
So both these terms can be rewritten to a common term.
\item (7) with (8): $0 + 1$ \\
$0 + 1 \rightarrow_{7} 1$ \\
$0 + 1 \rightarrow_{8} S(0) \rightarrow_{2} 1$ \\
So both these terms can be rewritten to a common term.
\item (6) with (9): $1 + 0$ \\
$1 + 0 \rightarrow_{6} 1$ \\
$1 + 0 \rightarrow_{9} S(0) \rightarrow_{2} 1$ \\
So both these terms can be rewritten to a common term.
\item (8) with (9): $1 + 1$ \\ 
$1 + 1 \rightarrow_{8} S(1)$ \\
$1 + 1 \rightarrow_{9} S(1)$ \\
So both these terms can be rewritten to a common term.
\item (10.0.0) with (1.0):  $0 \bzero{} + y \bzero{}$ as well as $x \bzero{} + 0 \bzero{}$ \\
$0 \bzero{} + y \bzero{} \rightarrow_{10.0.0} (\bzero{(0 + y)}) \rightarrow_{7} \bzero{y} $ \\
$0 \bzero{} + y \bzero{} \rightarrow_{1.0} 0 + \bzero{y} \rightarrow_{7} \bzero{y}$ \\
So both these terms can be rewritten to a common term. \\
$x \bzero{} + 0 \bzero{} \rightarrow_{10.0.0} \bzero{(x + 0)} \rightarrow_{6} \bzero{x}$ \\
$x \bzero{} + 0 \bzero{} \rightarrow_{1.0} \bzero{x} + 0 \rightarrow_{6} \bzero{x}$ \\
So both these terms can be rewritten to a common term.
\item (10.0.1) with (1.0): $0 \bzero{} + y \bone{}$ \\
$\bzero{0} + \bone{y} \rightarrow_{10.0.1} \bone{(0 + y)} \rightarrow_{7} \bone{y}$ \\
$\bzero{0} + \bone{y} \rightarrow_{1.0} 0 + \bone{y} \rightarrow_{7} \bone{y}$ \\
So both these terms can be rewritten to a common term.
\item (10.1.0) with (1.1): $\bone{0} + \bzero{y}$ \\
$\bone{0} + \bzero{y} \rightarrow_{10.1.0} \bone{(0+y)} \rightarrow_{7} \bone{y}$ \\
$\bone{0} + \bzero{y} \rightarrow_{1.1} 1 + \bzero{y} \rightarrow_{9} S(\bzero{y}) \rightarrow_{4} \bone{y}$ \\
So both these terms can be rewritten to a common term.
\item (10.1.1) with (1.1): $0 \bone{} + y \bone{}$ as well as $x \bone{} + 0 \bone{}$ \\
$\bone{0} + \bone{y} \rightarrow_{10.1.1} \bzero{S(0+y)} \rightarrow_{7} \bzero{S(y)} $ \\
$\bone{0} + \bone{y} \rightarrow_{1.1} 1 + \bone{y} \rightarrow_{9} S(\bone{y}) \rightarrow_{5} \bzero{S(y)}$ \\
So both these terms can be rewritten to a common term.

$\bone{x} + 0 \bone{} \rightarrow_{10.1.1} \bzero{S(x+0)} \rightarrow_{6} \bzero{S(x)}$ \\
$\bone{x} + \bone{0} \rightarrow_{1.1} \bone{x} + 1 \rightarrow_{8} S(\bone{x}) \rightarrow_{5} \bzero{S(x)}$ \\
So both these terms can be rewritten to a common term.
\item (13.0) with (1.0): $x \cdot 0 \bzero{}$ \\
$x \cdot \bzero{0} \rightarrow_{13.0} \bzero{(x\cdot 0)} \rightarrow_{11} \bzero{0} \rightarrow_{1.0} 0$ \\
$x \cdot \bzero{0} \rightarrow_{1.0} x \cdot 0 \rightarrow_{11} 0 $ \\
So both these terms can be rewritten to a common term.

\item (13.1) with (1.1): $x \cdot 0 \bone{}$ \\
$x \cdot \bone{0} \rightarrow_{13.1} \bzero{(x \cdot 0)} + x \rightarrow_{11} \bzero{0} + x \rightarrow_{1.0} 0 + x \rightarrow_{7} x $ \\
$x \cdot \bone{0} \rightarrow_{1.1} x \cdot 1 \rightarrow{12} x$ \\
So both these terms can be rewritten to a common term.
\end{enumerate}
\end{appendices}



\addcontentsline{toc}{section}{References}

\begin{thebibliography}{58}

\bibitem{baader1999term}
Baader, F. and Nipkow, T. (1999).
\emph{Term Rewriting and All That. }
\newblock Cambridge University Press.

\bibitem{bergstra2014three}
Bergstra, J.A. and Ponse, A. (2014). 
Three datatype defining rewrite systems for datatypes of Integers each extending a datatype of Naturals.
\newblock Preprint available: \url{arXiv/1406.3280v2} [cs.LO] (2014, 21 August).

\bibitem{kapur1990ground}
Kapur, D., Narendran, P. and Otto, F. (1990).
On ground-confluence of term rewriting systems.
\emph{Information and Computation}, 86(1):14-31.

\bibitem{knuth1983simple}
Knuth, D.E. and Bendix, P.B. (1983).
Simple word problems in universal algebras.
In \emph{Automation of Reasoning}, pages 342-376.
\newblock Springer.

\bibitem{dershowitz1979orderings}
Dershowitz, N. (1979).
Orderings for term-rewriting systems.
In \emph{Foundations of Computer Science, 20th Annual Symposium on Foundations of Computer Science}, pages 123-131.
\newblock IEEE.

\bibitem{specificatiedictaat}
Bethke, I. and Ponse, A. (2004).
Specificatietheorie (in Dutch).
\url{https://staff.fnwi.uva.nl/i.bethke/Specificatietheorie/dictaat.pdf}.

\bibitem{vgiesl2006aprove}
Giesl, J., Schneider-Kamp, P., and Thiemann, R. (2006). 
AProVE 1.2: Automatic termination proofs in the dependency pair framework. In 
U. Furbach and N. Shankar (Eds.): IJCAR 2006, 
Lecture Notes in Computer Science, Vol.~4130, Springer, pp.~281--286.

\bibitem{zankl2011csi}
Zankl, H., Felgenhauer, B., and Middeldorp, A. (2011). 
CSI - A confluence tool. 
\newblock In N.~Bj{\o}rner and V. Sofronie-Stokkermans (Eds.): CADE 2011, 
Lecture Notes in Computer Science, Vol.~6803, Springer,
pp. 499--505.

\bibitem{BW89}
Bouma, L.G. and  Walters, H.R. (1989).
Implementing algebraic specifications.
\newblock In J.A.~Berg\-stra, J.~Heering, and P.~Klint (Eds.): \emph{Algebraic Specification}
(Chapter 5), Addison-Wesley,
pp. 199--282.

\bibitem{walters1995rewrite}
Walters, H.R. and Zantema, H. (1995). 
Rewrite systems for integer arithmetic. 
\newblock In J.~Hsiang (Ed.): \emph{Rewriting Techniques and Applications 
(Proceedings 6th International Conference, RTA'95)}, 
Lecture Notes in Computer Science, Vol.~914, Springer, pp. 324--338.
Preprint available: \url{http://oai.cwi.nl/oai/asset/4930/4930D.pdf}.

\bibitem{bergstra2014fracpairs}
Bergstra, J.A. and Ponse, A. (2014). 
Fracpairs: fractions over a reduced commutative ring.
\newblock Preprint available: \url{arXiv/1411.4410v1} [math.RA] (2014, 17 November).

\end{thebibliography}
\end{document}